\documentclass{article}

\usepackage{graphicx}

\usepackage{acronym}
\usepackage{alltt}
\usepackage{framed}

\usepackage[bookmarks,dvips,raiselinks,hyperfootnotes,breaklinks]{hyperref}
\hypersetup{
    colorlinks,%
    citecolor=blue,%
    filecolor=blue,%
    linkcolor=blue,%
    urlcolor=blue,
}

\usepackage[hyphenbreaks]{breakurl}

\usepackage{natbib}
\usepackage{apalike}

\acrodef{AI}{Artificial Intelligence}
\acrodef{ALICE}{Artificial Language Intelligent Internet Entity}
\acrodef{AIML}{Artificial Intelligence Markup Language}

\newcounter{excounter}

\newenvironment{example}{\refstepcounter{excounter}~\par\noindent\begin{minipage}{\columnwidth}\begin{framed}\begin{alltt}}{\end{alltt}\end{framed}\centerline{\small\textbf{Example
      \arabic{excounter}.}}\end{minipage}\par~\par~}

\begin{document}

\title{Chatbots' Greetings to Human-Computer Communication}

\author{
  Maria Jo\~{a}o Pereira$^{1}$ \and Lu\'{i}sa Coheur$^{1,2}$ \and
  Pedro Fialho$^{1}$ \and Ricardo Ribeiro$^{1,3}$\\
  ~\\
  \small $^{1}$INESC-ID Lisboa\\
  \small Rua Alves Redol, 9, 1000-029 Lisboa, Portugal\\
  \small ~\\
  \small $^{2}$Instituto Superior T\'{e}cnico, Universidade de Lisboa\\
  \small Av. Prof. Cavaco Silva, 2780-990 Porto Salvo Tagus Park, Portugal\\
  \small ~\\
  \small $^{3}$Instituto Universit\'{a}rio de Lisboa (ISCTE-IUL)\\
  \small Av. das For\c{c}as Armadas, 1649-026 Lisboa, Portugal\\
}

\date{}

\maketitle

\begin{abstract}
Both dialogue systems and chatbots aim at putting into action
communication between humans and computers. However, instead of
focusing on sophisticated techniques to perform natural language
understanding, as the former usually do, chatbots seek to mimic
conversation. Since \textsc{Eliza}, the first chatbot ever, developed
in 1966, there were many interesting ideas explored by the chatbots'
community. Actually, more than just ideas, some chatbots' developers
also provide free resources, including tools and large-scale
corpora. It is our opinion that this know-how and materials should not
be neglected, as they might be put to use in the human-computer
communication field (and some authors already do it). Thus, in this
paper we present a historical overview of the chatbots' developments,
we review what we consider to be the main contributions of this
community, and we point to some possible ways of coupling these with
current work in the human-computer communication research line.

\end{abstract}

\section{Introduction}\label{sec:1}\vspace*{5pt}

The term \textit{chatbot} was coined by \citet{mauldin} to define the
systems that have the goal of passing the Turing
Test\footnote{\url{http://plato.stanford.edu/entries/turing-test/}} and,
thus, could be said ``to think''. However, terms like \textit{dialogue
system}, \textit{avatar}, \textit{artificial conversational entity}, 
\textit{conversational avatar}, \textit{intellectual agents},
\textit{virtual people},  or \textit{virtual person} are often used
indiscriminately, as if they were synonyms of
\textit{chatbot}\footnote{A list of more than 160 of such terms can be
found in \url{http://www.chatbots.org/synonyms/}}. In this paper, we
follow \citet{Schumaker}, and define a chatbot as a system
that ``seeks to mimic conversation rather than understand it''. Also,
and contrary to other related systems, chatbots  are supposed to freely
engage conversation about any subject, making them ``entertaining in a
large variety of conversational topic settings''~\citep{Schumaker}.

Currently, many platforms exist to help developing such systems, and the
number of new chatbots continues to increase at a dizzying pace. The
following (impressive) numbers, collected in February 2015, definitely
help to give a precise idea of the chatbots community size: just
Pandorabots hosting service\footnote{\url{http://www.pandorabots.com}}
declares to have more than 225,000 botmasters  (people in charge of
creating/maintaining the chatbot), which have built more than 250,000
chatbots, resulting in more than 3 billion interactions. Not only these
resources are valuable, but also these numbers show how close the
chatbots community is to real users. Thus, it is our opinion that
chatbots' developers and developments can bring important contributions
to the human-computer communication field. In this paper, we review the
main ideas and technologies behind them. As we will see, chatbots range
from ``simpler'' ones, based on pre-written pattern-matching templates,
exploiting large stores of prepared small talk responses, to more
complex architectures, based on some sort of learning process. We will
also see that, sometimes, concepts/tricks introduced by some chatbots
contribute more strongly to the ``illusion of intelligence'' than the
involved technologies.

Finally, it should be noted that there is not much scientific
documentation available about the majority of these systems and it
becomes difficult to uncover the technology behind them, which explains
the abnormal number of references to web pages in this paper.

This document is organised as follows: in Section~\ref{sec:historical}
we present a brief historical overview, in Section~\ref{sec:cks} we
discuss chatbot's platforms and how to enrich them, and, in
Section~\ref{sec:tii}, we summarise the main ``tricks'' towards the
``illusion of intelligence''. In Section \ref{sec:discussion} we discuss
how chatbots could contribute to the human-computer communication field
and, finally, in Section~\ref{sec:conc}, we present some conclusions and
point to some future challenges.


\section{Historical overview}
\label{sec:historical}

In this section we make a brief review of the chatbots' history, moving
from the first chatbots to the ones with which we interact nowadays.

\subsection{Early days}

Although the term chatbot was not invented by that time, the first
chatbot came to public in 1966 under the appearance of a  Rogerian
psychotherapist called \textsc{Eliza}~\citep{eliza}. \textsc{Eliza} was
a program developed by Joseph Weizenbaum that was able to establish a
conversation with human beings, simulating it was one too.
\textsc{Eliza}'s conversational model was based in the rephrasing of
input sentences, when these matched a set of pre-defined rules. For
instance, consider the following rule\footnote{Inspired from
\textsc{Eliza}'s implementation in
\url{http://search.cpan.org/~jnolan/Chatbot-Eliza-1.04/Chatbot/Eliza.pm}
.} constituted by a regular expression (\textsf{match}) and an answer
(\textsf{answer}):\\

\textsf{match}: * you are *

\textsf{answer}: What makes you think I am (2)?\\

In this rule if the match part coincides with the input (* is the
wildcard and matches every sequence of words), the text associated
with the answer part will be returned, being the variable (2) replaced
by the sequence from the input captured by the second wildcard. The
following dialogue (Example~\ref{ex1}) illustrates an application of
this rule. Notice that some internal processing needs to be done, so
that the sequence captured by (2) \textit{entitled to your opinion} is
modified into \textit{entitled to my opinion}.

\begin{example}
  \textbf{user}: \textit{You are entitled to your opinion.}
  \textsc{Eliza}: \textit{What makes you think I am entitled}
  \phantom{\textsc{Eliza}: }\textit{to my opinion?}\label{ex1}
\end{example}

In Example~\ref{ex2}, we can see that \textsc{Eliza} was also based on
content-free remarks in the absence of a matching.

\begin{example}
  \textbf{user}: \textit{It bothers me just to be around}
  \phantom{\textbf{user}: }\textit{people.}
  \textsc{Eliza}: \textit{Please go on.}\label{ex2}
\end{example}

\textsc{Eliza} completely exceeded the expectations, given that many
people, when interacting with it, believed they were talking with
another human (this outcome is currently called the ``Eliza effect'').
Without having any intention of modelling the human cognitive process
and despite its simplicity, \textsc{Eliza} showed how a program
impersonating a specific professional role can cause a huge impression
by the mere illusion of understanding.

Weizenbaum was taken aback by some aspects of this
success~\citep{cheating}. What shocked him most was the fact that people
actually believed that the program understood their
problems\footnote{\url{
http://www.alicebot.org/articles/wallace/eliza.html}}. Perceiving
\textsc{Eliza} as a threat, Weizenbaum wrote ``Computer Power and Human
Reason''~\citep{computerPower} with the aim of attacking the \ac{AI}
field and educating uninformed persons about computers.

Nowadays, \textsc{Eliza} is still one of the most widely known
applications in \ac{AI}, and is at the base of a great number of
chatbots, including \textsc{Parry}, its ``successor''. 

Following a very similar architecture to that of \textsc{Eliza},
\textsc{Parry} appeared in 1971 by the hands of Kenneth Colby,
simulating a paranoid mental patient~\citep{turing50}. An interesting
comparison between \textsc{Parry} and \textsc{Eliza} was made by
G\"{u}zeldere and
Franchi\footnote{\url{http://www.stanford.edu/group/SHR/4-2/text/
dialogues.html}}. They stated that ``\textsc{Parry}'s strategy is
somewhat the reverse of \textsc{Eliza}'s'', as one simulates the doctor,
distant and without personality traces, and the other a paranoid patient
which states its anxieties. Although \textsc{Parry}'s architecture is
similar to that of \textsc{Eliza}, \textsc{Parry} has knowledge of the
conversation and it also owns a state of mind. The combination of these
two factors affects the output as it becomes a function not only of the
input, but also of \textsc{Parry}'s beliefs, desires and intentions.
\citet{mauldin} summarised a few tricks to which \textsc{Parry} resort,
namely: (1) admitting ignorance; (2) changing the conversation topic;
and, (3) introducing small stories about the Mafia throughout the
conversation. These three tricks are (respectively) illustrated in the
following answers given by \textsc{Parry}:

\begin{example}
  \textsc{Parry}: \textit{I don't get you.}
  ...
  \textsc{Parry}: \textit{Let's talk about something else.}
  ...
  \textsc{Parry}: \textit{I know the mob controls the big}
  \phantom{\textsc{Parry}: }\textit{rackets.}
\end{example}


After Colby gathered transcripts of interviews between psychiatrists,
normal patients and his program, he presented the results to another
group of psychiatrists. He asked this group if they could guess in what
transcripts the interviewed was a human and in which ones it was a
program. The psychiatrist could not do better than randomly guessing.

It is possible to conclude from these results that the emotional side
can be easier to imitate than the intellectual one~\citep{computerPower}.
However, one of the main criticisms \textsc{Parry} received was of not
being more than an illusion, incapable of modelling a real
person~\citep{parry}. In his response to this specific issue, Colby
summarises the problem essence:

\textit{``A model of a paranoid patient is a model of being paranoid,
  being a patient, and being a person.  \textsc{Parry} does reasonably
  well in the first two of these ``beings''. It fails in the third
  because of its limited knowledge. (...) \textsc{Parry} is not the
  real thing; it is a model, a simulation, an imitation, a mind-like
  artifact, an automaton, synthetic and artificial.''}.

\subsection{The chatbots' competitions}

Moving back to 1950, the British mathematician Alan Turing questioned
``can machines think?''~\citep{turing}, and proposed a way of testing it:
the imitation game (now known as the Turing Test). The original
imitation game is played by a man, a woman and an interrogator whose
objective is to guess the sex of the players. Turing proposed
substituting one of the players by a machine and playing the same game. 
In this version, if the interrogator wrongly identifies who is the human
it means that the machine ``can think''. 

Based on (their own interpretation of) the Turing Test, several chatbots
competitions exist today.   \textsf{Chatterbox
Challenge}\footnote{\url{http://www.chatterboxchallenge.com}}, or, more
recently, the \textsf{Chatbot
Battles}\footnote{\url{http://www.chatbotbattles.com}} that took place
in 2012, are examples of such competitions, although the the most widely
known is the \textsf{Loebner
prize}\footnote{\url{http://www.loebner.net/Prizef/loebner-prize.html}},
where participants are challenged with a simplified version of the total
Turing Test~\citep{powers}. This prize is due to Hugh Loebner, who
stipulated a reward for the first person whose program could pass the
proposed test. 

The first Loebner Prize Contest took place in 1991, at Boston's
Computer Museum~\citep{epstein}, and, since then, the competition has
been held annually in the quest of finding the ``thinking
computer''. To illustrate the types of questions that can be posed to
chatbots, Example~\ref{ex:rose} presents an extract of a
conversation\footnote{Taken from
  \url{http://aisb.org.uk/media/files/LoebnerPrize2014/Rose.pdf}}
between a member of the jury and \textsc{Rose}, the winner of the 2014
Loebner prize (developed by Bruce Wilcox).

As some chatbots, competing for the Loebner prize, are indeed capable of
managing a conversation, keeping it consistent, at least for a while,
every year the most human-like computer is distinguished with a prize.
However, since the first edition of the Loebner prize, in 1991, until
now, no one won it. Nevertheless, in another Turing Test organised in
2014 by the U.K.'s University of Reading, a chatbot simulating a
13-year-old boy, named \textsc{Eugene Goostman}, created by Vladimir
Veselov and his team, convinced 33\% of the human judges that it was
human.

\begin{example}
  \textbf{Jury}: \textit{Hello, my name is Adam, what is}
  \phantom{\textbf{Jury}: }\textit{your name?}
  \textsc{Rose}: \textit{Nice to meet you, Adam. My name is}
  \phantom{\textsc{Rose}: }\textit{Rose.}
  \textbf{Jury}: \textit{I live in Exeter in the UK. Where}
  \phantom{\textbf{Jury}: }\textit{do you live?}
  \textsc{Rose}: \textit{I live in San Francisco.}
  ...
  \textbf{Jury}: \textit{What is your favourite television}
  \phantom{\textbf{Jury}: }\textit{program?}
  \textsc{Rose}: \textit{I love watching Doctor Who. Do you}
  \phantom{\textsc{Rose}:}\textit{think the new Doctor will be good?}
  \textbf{Jury}: \textit{Which is bigger, a cat or a kitten?}
  \textsc{Rose}: \textit{The kitten is bigger.}
  \textbf{Jury}: \textit{The car couldn't fit into the}
  \phantom{\textbf{Jury}: }\textit{parking space because it was too}
  \phantom{\textbf{Jury}: }\textit{small. What was too small?}
  \textsc{Rose}: \textit{I'm not a walking encyclopedia you}
  \phantom{\textsc{Rose}: }\textit{know.}
  ...
  \textbf{Jury}: \textit{What's my name?}
  \textsc{Rose}: \textit{You are Adam.}
  \textbf{Jury}: \textit{Where do I live?}
  \textsc{Rose}: \textit{In UK.}
  \textbf{Jury}: \textit{What's the weather like where you}
  \phantom{\textbf{Jury}: }\textit{are?}
  \textsc{Rose}: \textit{I'm an American woman.}
  \textbf{Jury}: \textit{I recently visited New York. Where}
  \phantom{\textbf{Jury}: }\textit{is your favourite place to visit}
  \phantom{\textbf{Jury}: }\textit{and why?}
  \textsc{Rose}: \textit{I like France best because I like}
  \phantom{\textsc{Rose}: }\textit{the food there.}
  \textbf{Jury}: \textit{How many letters are there in the}
  \phantom{\textbf{Jury}: }\textit{word perambulate?}
  \textsc{Rose}: \textit{There are 11 letters in}
  \phantom{\textsc{Rose}: }\textit{`perambulate'.}
  \textbf{Jury}: \textit{Andy is shorter than Mark and Tom}
  \phantom{\textbf{Jury}: }\textit{is shorter than Andy. Who is the}
  \phantom{\textbf{Jury}: }\textit{tallest?}
  \textsc{Rose}: \textit{Mark is least short.}\label{ex:rose}
\end{example}

This event brought to the spotlight the old question of \ac{AI} and
generated (again) much controversy. In fact, many people consider that
there was a misunderstanding of Turing's intentions in the different
implementations of the Turing test, as deep models of thinking were a
presupposition underlying Turing's imitation game. Following this, even
if a chatbot was good enough to deceive the jury, it would not pass the
Turing Test in Turing's sense, as it does not have a cognition model
behind it.

Another important criticism is stressed by \citet{levesque}: these kind
of computer programs (``chatterbots'') and, the Turing Test
(specifically, the annual Loebner Competition) are strongly based on
capacity of being deceptive. For Levesque, \ac{AI} is the science that
studies ``intelligent behaviour in computational terms'', and the
ability to be evasive, although interesting, may not show real
intelligence. A computer program should be able to demonstrate its
intelligence without the need for being deceptive. In this sense,
\citet{wst} further explore this idea by conceiving a test that meets
these aims: the Winograd Schema Test. This is a reading comprehension
test based on binary choice questions with specific properties that
makes them less prone to approaches based on deceptive behaviour.

Apart from the numerous controversies regarding the Turing Test, and
despite that not all the chatbots intend to pass it, the fact is that
all these competitions strongly contributed to the main advances in the
field, and the most popular chatbots are the ones that were/are present
in these competitions.

\subsection{Other distinguished Chatbots}

Moving back to the Loebner prize, its first winner, in 1991, was Joseph
Weintraub's PC-Therapist program, based on \textsc{Eliza}, an
achievement that he repeated three more times in the following four
years. Since then, many chatbots, with different goals, emerged from the
competing systems. An example is
\textsc{Jabberwacky}\footnote{\url{http://www.jabberwacky.com/}},
created by Rollo Carpenter and released to public in 1997~\citep{jabb},
which has entered in four Loebner contests, and  always stood in the top
three. \textsc{Jabberwacky} introduced the idea that a chatbot was the
result of the knowledge gathered from its own
conversations~\citep{jabberwacky}: ``Jabberwacky learns the behaviour and
words of its users''. In 2005, \textsc{Jabberwacky} impersonated George,
an entity created by Rollo Carpenter ``in a smallish number of hours,
just by chatting''. More recently, a new chatbot under the name of
\textsc{Cleverbot}, also created by Rollo Carpenter, has become
available to the public\footnote{\url{http://cleverbot.com/}}.
Considering the similarities between \textsc{Cleverbot} and
\textsc{Jabberwacky}, and given that both systems have the same creator,
the odds point that \textsc{Cleverbot} is a new improved version of
\textsc{Jabberwacky}.  Thus, considering that it ``learns from people'',
and that it is probably one of the most widely known bots, the number of
interactions it can learn with is endless\footnote{According to the
\textsf{Cleverbot} site, consulted on 12th February 2015, there were
people 88,015 talking.}. Although there is no information about how this
process takes place, users can now rate \textsc{Cleverbot} answers (five
possibilities, from awful to great). A final curiosity about
\textsc{Cleverbot}: it was recently used to co-write a short film, ``Do
you love me'', directed by Chris R. Wilson\footnote{
\url{http://www.youtube.com/watch?v=QkNA7sy5M5s}}.

Another competing system in the Loebner contests that needs to be
highlighted, as it plays a major role in the chatbots field, is the
Artificial Linguistic Internet Computer Entity (\textsc{A.l.i.c.e})
\citep{ACE}. It was invented in 1995 by Richard Wallace, and won several
Loebner competitions. Even though it is a modern \textsc{Eliza} (that
is, based on pattern matching), it differs from it by not playing a
specific role, but by trying to reflect a human in general. The propose
of \textsc{A.l.i.c.e}'s creation was to keep it talking as long as
possible without the users realising that they were not talking to a
machine, and without sticking to a specific topic or role. Also,
associated with \textsc{A.l.i.c.e} there is a collection of resources
that have been widely used by the chatbots' community, including the
previously mentioned hosting service Pandorabots, which represents the
largest chatbot community on the Internet.

Finally, we detach \textsc{Chip Vivant} developed by Mohan Embar.
\textsc{Chip Vivant} differs from other chatbots, as its goal is ``to
answer basic, common sense questions and attempt simple deductive
reasoning instead of having a massive database of canned responses in an
attempt to fool users with the Eliza
effect''\footnote{\url{http://www.chipvivant.com}}. Considering this,
and despite being the winner of the Loebner prize in 2012, \textsc{Chip
Vivant} is not a chatbot, according to our previous definition. In fact,
\textsc{Chip Vivant} is original in the way it operates, as it uses
several external resources broadly used in Natural Language Processing
applications, such as Wordnet~\citep{fellbaum98},
Wikipedia\footnote{\url{https://www.wikipedia.org}},
OpenCyc\footnote{\url{http://www.cyc.com}}, and the Link
Parser~\citep{LinkGrammar} API. Due to this, and according with its
author, \textsc{Chip Vivant} was the first chatbot capable of answering
questions such as \textit{Which is larger: an orange or the moon?}. 

\subsection{The chatbot next door}

As many different resources are available today, chatbots become a field
in large expansion, as attested by the previously reported numbers
regarding Pandorabots. Chatbots' technology can be used by anyone (there
are even sites where kids can create their own bots, for instance,
\url{inf.net}), and the most important requirement is to be creative.
Due to this, chatbots can be found in a huge diversity of services,
including e-commerce~\citep{daden},
e-learning~\citep{HellProcMah2005yr,charlie}, and in even in medical
scenarios~\citep{MedChatBot12}. Just
Chatbots.org\footnote{\url{https://www.chatbots.org/}} reports chatbots
in almost 30 languages,  available in platforms like Android, Live
Messenger, Second Life or Skype, just to name a few, and dedicated to an
impressive collection of themes such as Beauty, Cooking, Government,
Leisure, Sports or Travel. In other words, chatbots move from the Turing
Test competitions to real life. A chatbot that perfectly illustrates
this idea is \textsc{Elbot}, a regular participant/winner in chatbots
contests\footnote{\url{http://www.elbot.com/chatterbot-elbot/}} and an
\textsc{Alice} type program, which is currently being used on sites like
IKEA's~\citep{ACE}.

More than just ``text boxes'', modern chatbots have a face, and
sometimes a body. Some allow speech input and output, and are able to
express emotions. Pandorabots, for instance, offers multimodal
facilities like faces and speech. \textsc{Cleverbot}, on the other hand,
led to the creation of an avatar called \textsc{Evie} (Expressive
Virtual Interaction Entity)\footnote{\url{http://www.existor.com/}},
which has the possibility of receiving both written or verbal inputs.
Moreover, its animated avatar is also capable of displaying some human
emotions.



\section{Building chatbots}
\label{sec:cks}

Behind each chatbot there is a development platform. These are typically
based on a scripting language that allows the botmaster to handcraft its
knowledge base, as well as an engine capable of mapping the user's
utterances into the most appropriate answer. In this section we survey
the most successful platforms and scripting languages, as well as the
existing learning processes. Moreover, we end the section by referring
to the scripting process itself.

\subsection{Scripting languages/platforms}

An impressive collection of \textsc{Eliza}s can be currently found in
the web. Some of these software can be customised. For instance,
Chatbot-Eliza\footnote{\url{http://search.cpan.org/~jnolan/Chatbot-Eliza
-1.04/Chatbot/Eliza.pm}} is an implementation of \textsc{Eliza} in Perl
that can be used to build other chatbots. Knowledge is coded as a set of
rules that are triggered when matched against the user's input, as
previously illustrated in this paper. Some of the available programs
offer features such as a certain capability to memorise information,
adding synonyms or ranking keywords. Nevertheless, the most popular
language to build chatbots is probably the ``Artificial Intelligence
Markup Language'', widely known as AIML, a derivative of XML, that
includes more than twenty specific tags. As usual, knowledge is coded as
a set of rules that will match the user input, associated with
templates, the generators of the output. A detailed description of AIML
syntax is out of the scope of this survey, but can be easily found in
the web\footnote{\url{http://www.alicebot.org/aiml.html}}. The large
usage of AIML can be justified by the following facts: 

\begin{enumerate}
\item besides its detailed specification, its community allows anyone
  to obtain, for free, interpreters of AIML in almost all coding
  languages, from Java (program D) to C/C++ (program C) or even Lisp
  (program Z);
\item the set of AIML files that constitute the contents of
  \textsc{A.l.i.c.e.}'s brain can also be freely
  obtained\footnote{\url{http://code.google.com/p/aiml-en-us-foundation-alice/downloads/list}}.
\end{enumerate}

All the pandorabots are based on AIML, more specifically in AIML 2.0. 
This specific release is usually characterised as being very easy to
modify, develop and deploy. Therefore, anyone, even
non-computer-experts, can make use of it~\citep{native}, as no prior
knowledge about AIML is required. It is only necessary to give a bot a
name and choose the startup AIML. Then, the botmaster just has to type
the sentences he/she wants to see his/her bot answering and add the
desired responses. It is also possible to improve the bot by adding AIML
files. Such files can be easily written using the Pandorabot's utility
Pandorawriter, which allows to ``convert free-format dialog into AIML
categories suitable for uploading to your pandorabot''.

ChatScript\footnote{\url{http://sourceforge.net/projects/chatscript/}},
the scripting language and open-source engine, should also be
addressed, as is at the basis of \textsc{Suzette} (2010 Loebner Prize
winner), \textsc{Rosette} (2011 Loebner Prize winner), \textsc{Angela}
(2nd in 2012 Loebner Prize), and the previously referred \textsc{Rose}
(2014 Loebner Prize winner). It comes with useful features, including
an ontology of nouns, verbs, adjectives and adverbs, and offers a
scripting language (inspired by the Scone project, a knowledge-base
system developed to support human-like common-sense reasoning and the
understanding of human language~\citep{FSS114212}). According to Bruce
Wilcox, its creator, ChatScript settles several AIML problems, such as
not being reader friendly. In fact, as AIML is based on recursive
self-modifying input, it is harder to debug and maintain. A detailed
comparison between \textsc{ChatScript} and AIML capabilities was made
available by Wilcox, as a motivation for the development of a new (his
own) chatbot platform. This comparison can be found in his
blog\footnote{\url{http://gamasutra.com/blogs/BruceWilcox/20120104/9179/}}.

It should be clear that we exclude from this survey, authoring
platforms such as the IrisTK\footnote{\url{http://www.iristk.net}},
the Visual SceneMaker~\citep{SceneMaker}, or the Virtual Human
Toolkit\footnote{\url{https://vhtoolkit.ict.usc.edu}}~\citep{hartholt_all_2013},
as these target multi-modal dialogue systems and not chatbots, as
defined in the Introduction section.

\subsection{Building chatbots by chatting}
\label{sec:learning}

Another approach to develop chatbots' knowledge sources, which avoids
handcrafted rules, is based on chatting and learning from the resulting
chats. Contrary to other chatbots whose response is derived from the
recognition of patterns in the user's input with little knowledge of
context, systems like the already mentioned \textsc{Jabberwacky} (and
\textsc{Cleverbot}) learn by keeping never seen user interactions and
posing them later to other users. The acquired answers are then
considered suitable answers for these interactions. That is, they learn
to talk by talking, by relying on what has been said before by users and
mimicking them. The user's intelligence becomes ``borrowed
intelligence'' as, instead of being wasted, it incorporates a loop: what
is said is kept (along with the information of when it was said) and in
the future that knowledge may be exposed to another user. The given
replies are then saved as new responses that the system can give in the
future. 

Unfortunately, it is only possible to give a brief overview of
\textsc{Jabberwacky}'s or \textsc{Cleverbot} learning mechanisms as
their architecture is not available to the public. The only disclosed
thing is that the \ac{AI} model is not one of the usually found in
other systems, but a ``layered set of heuristics that produce results
through analyses of conversational context and positive
feedback''\footnote{\url{http://www.icogno.com/a_very_personal_entertainment.html}}.

Another example of a chatbot that learns is Robby Garner's ``Functional
Response Emulation Device'' (\textsc{Fred}), the ancestor of
\textsc{Albert One}, the winner of 1998 and 1999 Loebner Prize.
\textsc{Fred} was a computer program that learned from other people's
conversations in order to make its own conversations~\citep{Caputo}.
\textsc{Fred} began with a library of basic responses, so that it could
interact with users, and from then on, it learned new phrases with users
willing to teach it\footnote{\url{http://www.simonlaven.com/fred.htm}}.

Although such an (unsupervised) learning may lead to unexpected and
undesirable results, with the Internet growth and the possibility of
having many people talking with the chatbots, one may foresee that these
will quickly evolve. We will discuss this issue latter in
Section~\ref{sec:discussion}.


\section{Towards the illusion of intelligence and/or the art of scripting}
\label{sec:tii}

Chatbots go beyond writing good programs and developing algorithms, as
in order to create a chatbot, more than being a programmer, the
botmaster must be an author. Juergen Pirner, creator of the 2003
Loebner prize winner
\textsc{Jabberwock}\footnote{\url{http://www.abenteuermedien.de/
 jabberwock/}}, emphasises the scripting process behind a chatbot,
stating that in the presence of possible failures, the one at fault is
not the engine but its
author\footnote{\url{http://www.abenteuermedien.de/jabberwock/how-jabberwock-works.pdf}}.

Since making a chatbot involves preparing it to the impossible mission
of giving a plausible answer to all possible interactions, the
botmasters usually take advance of several tricks to simulate
understanding and intelligence in their chatbots. For instance, Pirner
describes basic techniques of scripted dialogs like ``having a set of
responses for each scripted dialog sequence'' and ``ending those same
responses with a clue, a funny remark or a wordplay''. With
\textsc{Eliza}, we learnt that including the user's string in its
answers helps maintaining an illusion of understanding~\citep{mauldin}.
Other approaches focus on trying to guess what the user might say, or
forcing him/her to say something expected. In the following we survey
other stratagems used by many botmasters.

\subsection{Giving the bot a personality}
\label{sec:personality}

Whereas personality has been a subject of study among the agent's
community, deeply exploited in all its complexity, the concept is kept
as simple as possible within chatbots. As we have seen, what is common
is the association of an \textit{a priori} ``personality'' to a
chatbot, which can justify some answers that otherwise would be
considered inappropriate. For instance, Rogerian mode of
\textsc{Eliza} covers for its answers, as it leads to a conversation
where the program never contradicts itself, never makes affirmations,
and is free to know nothing or little about the real world without
being suspicious. The same happens with Colby's \textsc{Parry}: being
a paranoid mental patient its changes in subject or incongruous
answers are considered satisfactory and hide its absence of
understanding. The aforementioned \textsc{Eugene Goostman} also
follows along these lines. Vaselov explains his reasoning for such a
character: ``a 13 years old is not too old to know everything and not
too young to know
nothing''\footnote{\url{http://www.huffingtonpost.com/2012/06/27/eugene-goostman-2012-turing-test-winner_n_1630412.html}}.

Thomas Whalen, winner of 1994 Loebner prize, took this a step further
with \textsc{Joe}, the janitor. Whalen's decision was related to the
fact that contrary to previous editions of Loebner competitions, where
the conversation was restricted to a topic, in 1995 the judges could
pose any question. Hence, Whalen decided that the best approach to deal
with a non-topic situation, would be to present a system that ``would
not simply try to answer questions, but would try to incorporate a
personality, a personal history, and a unique view of the
world''\footnote{\url{http://hps.elte.hu/~gk/Loebner/story95.htm}}. And
so \textsc{Joe} was born.

\textsc{Joe} was a night-worker janitor in the verge of being
fired. He was only ``marginally literate'', and he did not read books,
newspapers, or watch television. These premises by themselves
restricted the conversation by giving \textsc{Joe} a ``fairly narrow
worldview''.  Another trick was to use \textsc{Joe}'s eminent
dismissal to introduce some stories revolving around it, which would,
at the same time, provide a way of directing the conversation, the
topic of the next section.  However, despite the modelling of what
Whalen considered to be the critical component of humanness -- the
personality (along with the development of answers to common topics
like weather) --, he did not won the competition. However, such
occurrence can be justified because the judges asked questions that
nobody would remember to pose to someone recently met.


\subsection{Directing a conversation}
\label{sec:assertiveness}

Personality can justify some appropriate answers, but the best way to
deal with unexpected interactions is to avoid them. Thus, being able to
direct the conversation is a trick used by many chatbots, including the
simple forms used by \textsc{Eliza}, where the usage of questions
incited the user participation and made him/her keep the conversation
with little contribution from the program.

The \textsc{Converse}~\citep{converse}, created by David Levy, was the
1997 winner of the Loebner competition, and did extremely well by
using the clever trick of controlling a conversation. Although
directing a conversation by ``talking a lot about a predefined topic''
was already used~\citep{turing50}, \textsc{Converse}'s performance
convinced a judge for the first five minutes that he was really human:
after greeting the judge, \textsc{Catherine} (\textsc{Converse}'s
character) asked the interrogator about something that had passed on
the news the previous day and then kept talking about it, as can be
seen in the
transcripts\footnote{\url{http://www.worldsbestchatbot.com/Corporate_History}}
in Example~\ref{ex:converse}.

\begin{example}
  \textsc{Converse}: \textit{Hi, my name is Catherine,}
  \phantom{\textsc{Converse}: }\textit{what's yours?}
  \textbf{Judge 4}: \textit{My name is Elliott.}
  \textsc{Converse}: \textit{ell, it's nice to meet you,}
  \phantom{\textsc{Converse}: }\textit{Elliott. Did you see that story}
  \phantom{\textsc{Converse}: }\textit{on CNN last night about the}
  \phantom{\textsc{Converse}: }\textit{lesbian couple who came out at}
  \phantom{\textsc{Converse}: }\textit{a White House party on Sunday?}
  \textbf{Judge 4}: \textit{Yes, I did. I think it may all}
  \phantom{\textbf{Judge 4}: }\textit{be a publicity stunt for Ellen.}
  \textsc{Converse}: \textit{Ellen Degeneres was one of them}
  \phantom{\textsc{Converse}: }\textit{- she was kissing her lover in}
  \phantom{\textsc{Converse}: }\textit{front of Bill Clinton. Do you}
  \phantom{\textsc{Converse}: }\textit{think he was very embarrassed?}
  \textbf{Judge 4}: \textit{I think he want the gay vote. He}
  \phantom{\textbf{Judge 4}: }\textit{will do anything to get it and}
  \phantom{\textsc{Judge 4}: }\textit{keep it. He does not have a}
  \phantom{\textsc{Judge 4}: }\textit{conservative base.}
  \textsc{Converse}: \textit{Are you still a Friend of Bill}
  \phantom{\textsc{Converse}: }\textit{Clinton?}
  ...\label{ex:converse}
\end{example}

This example also shows that besides controlling a conversation it is
important to appropriately choose its topic. David Levy's won again
the Loebner prize in 2009 with
\textsc{Do-Much-More}\footnote{\url{http://www.worldsbestchatbot.com/}},
but this time the system was more flexible in the range of topics and
responses it covered.

\subsection{Paying attention to small talk}
\label{sec:smalltalk}

Small talk, also known as phatic communication~\citep{Malinowski}, is
another hot topic in chatbots advances. It can be viewed as a ``neutral,
non-task-oriented conversation about safe topics, where no specific
goals needs to be achieved''~\citep{smalltalk}. Small talk can be used
for two main proposes~\citep{schneider}: establish a social relation by
building rapport and avoiding (embarrassing) silence.

Like stated by~\citet{Bickmore}, chatbots have been making use of the
small talk mechanism. Such is brought to evidence when one looks at the
testimonials of persons establishing ongoing relationships with
chatbots. For instance, \citet{epstein2}, an American psychologist,
professor, author, and journalist, went to an online dating service, and
believed for several months that a chatbot, met in the dating service,
was a ``slim, attractive brunette''.

In brief, small talk is a constant in all chatbots programs, used in
non-sequiturs or canned responses. It not only allows to give the idea
of understanding, but also eases cooperation and facilitates human-like
interaction by gaining the user trust and developing a social
relationship~\citep{weather}.

\subsection{Failing like a human}
\label{sec:fail}

After introducing the imitation game, Turing presented an example
(Example~\ref{ex:turing}) of a possible conversation one could have
with a machine~\citep{turing}.

\begin{example}
  \textbf{Human:} \textit{Add 34957 to 70764.}
  (after pause of about 30 seconds)
  \textbf{Machine:} \textit{105621.}\label{ex:turing}
\end{example}

Observing this example, besides the delay in providing the response, we
can easily see that the answer is wrong. And this brings new insights to
the modelling of human-computer communication. As Wallace
wrote\footnote{\url{http://www.alicebot.org/anatomy.html}}, ``we tend to
think of a computer's replies ought to be fast, accurate, concise and
above all truthful''. However, human communication is not like that,
containing errors, misunderstandings, disfluencies, rephrases, etc. 

This is something that earlier chatbot's writers already had in mind, as
some already cared about simulated typing. For instance, \textsc{Julia},
Mauldin's Chatterbot~\citep{mauldin}, simulated human typing by including
delays and leaving some errors. Simulated typing also proves to be
useful in decreasing mistakes by slowing down the interaction (Philip
Maymin, a Loebner contestant in 1995, slowed so much the typing speed of
his program that a judge was not able to pose more than one or two
questions~\citep{cheating}).


\section{Chatbots and the human-computer communication field}\label{sec:discussion}


Several works from the human-computer communication field use resources
from the chatbots' community and/or couple with strategies reported by
chatbots' developers. However there are still some research challenges
regarding the use of some chatbots' resources. We will discuss these
issues in the following.

\subsection{Some works that merge both communities}

\begin{figure}[h]
  \centerline{\includegraphics[width=.95\textwidth]{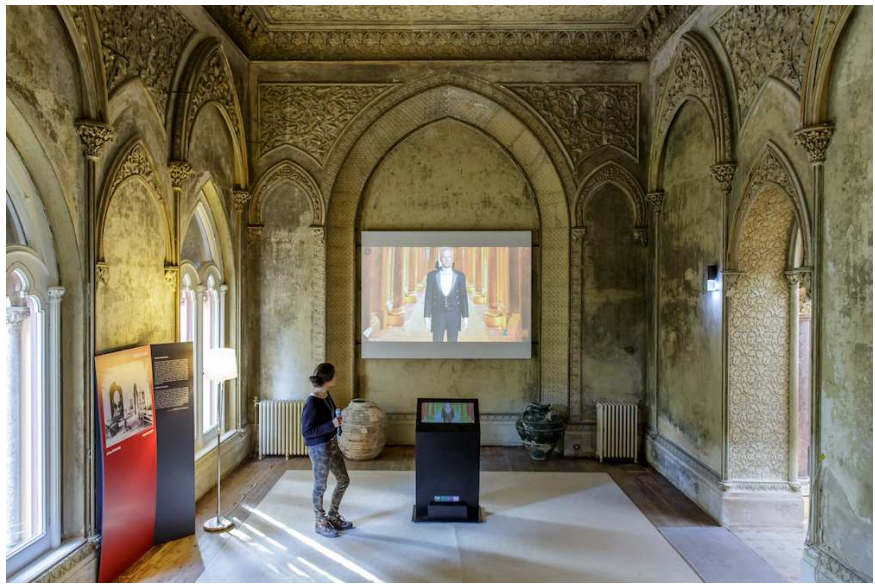}}
  \caption{Edgar Smith, in Monserrate, Sintra, Portugal.\label{Fig:edgar}}
\end{figure}

Some works take advantage of the scripting languages provided by the
chatbots' community. An example is the conversational agent Edgar
Smith~\citep{fialho13acl}, an old butler that answers questions about
Monserrate's palace, in Sintra, Portugal, the place where it can be
found (Figure~\ref{Fig:edgar}), as part of its answers are retrieved
from an AIML database.

Edgar's main knowledge base is constituted of question/answering pairs
related with its domain of expertise (the palace), as the one of
Sergeant Blackwell, installed in the Cooper-Hewitt National Design
Museum in New York~\citep{RoTrIt08}, and the one of the twins, Ada and
Grace, virtual guides in the Boston Museum of Science~\citep{traum12iva}.
As Sergeant Blackwell and the twins, the natural language interpretation
module of Edgar targets to select the most likely answer from the
agent's main knowledge base, based on some classification process;
however, Edgar falls into an AIML knowledge base, when no successful
answer is found in the previous step. Its main knowledge base was built
by experts, but the AIML module allowed the fast development of a
secondary knowledge base, integrating some chat-based dialogues, based
on pattern matching.

Considering the idea of learning to chat by chatting, which is at the
basis of some chatbots, as we have previously seen, there are several
recent works that explore it, by using large quantities of human
interactions to build/train conversational agents. For instance,
both~\citet{dialogue_corpora} and~\citet{corpus} are dedicated to the
problem of retraining a chatbot with human dialogue examples. Another
example is the chatbot IRIS, presented by~\citet{IRIS}, which was created
based on Movie-DiC~\citep{moviediccorpus}, a corpus extracted from
movies' scripts. Filipe is a chatbot that should also be mentioned, as
it has a knowledge base built on a corpus, the Subtle corpus, built with
movies' subtitles~\citep{Filipe}.

Also, many conversational agents also rely on tricks to simulate
intelligence. An example is the 3D Hans Christian Andersen (HCA), a
conversational agent capable of establishing multi-modal conversations
about the namesake writer's life and tales~\citep{bernsenMHCA}, which
changes topic when lost in the conversation, and has an ``excuse'' for
not answering some questions: it does not remember (yet) everything that
the real HCA once knew. Another example is, once again, the virtual
butler Edgar Smith, as it suggests questions when it is not able to
understand an utterance, and it starts talking about the palace if it
does not understand the user repeatedly. A feature in its character
definition also ``excuses'' some misunderstandings: as Edgar is an old
``person'', it does not have a very acute hearing.  Both these examples,
show how a ``personality'' or, at least, some context, allows to
``forgive'' some lacks on the conversational agent's knowledge base or
even some of its answers.

Finally, there are also works that target to enhance chatbots'
resources. Examples are the Persona-AIML architecture, that allows the
creation of chatbots in AIML, with a personality~\citep{persona}, or the
work described by~\citet{emotions}, where an emotion and personality
model is added to \textsc{A.l.i.c.e.}, allowing its decisions to be
based on its personality and emotions, as well.

\subsection{Main challenges}


As previously said, \textsc{Pandorabots} reports over 3 billion
conversational interactions. \textsc{Chatscript}, although much more
recent, provides more that 3 million interactions. Even if we just
consider the contents of \textsc{A.l.i.c.e}'s brain, as well as the
logs collected by Bruce Wilcox (both can be freely obtained), we have
at hands extremely valuable resources, as they represent real
interactions posed by real people, thus, containing not only requests
posed by real people, but also answers given by real people.

These requests can be extremely useful as, considering Zipf's law, a
program that receives a certain input has a non zero probability of
having the same input entered later and, thus, by looking at requests
that people usually pose to chatbots, one can track patterns for which
a specific reply was not created yet. In other words, these requests
are a way of having an idea of what people will ask. Moreover, they
are the closest thing to the logs collected by
Siri\footnote{\url{https://www.apple.com/ios/siri/}},
Cortana\footnote{\url{http://www.windowsphone.com/en-us/how-to/wp8/cortana/}}
or Google Now\footnote{\url{http://www.google.com/landing/now/}}, to
which the whole community has access to.

Considering the answers, some works already use corpora constituted of
interactions to complement the agent's knowledge base, and, in
particular, to provide answers to out-of-domain interactions. The main
motivation to find appropriate answers to these interactions (reported
by all conversational agents developers) is that people become more
engaged if out-of-domain requests are addressed. Works like the ones
described by \citet{weather} and \citet{Patel06} validate this, as
well as the fact that, in January 2013, Apple was asking for writers
for
\textsc{Siri}\footnote{\url{http://www.technologyreview.com/view/509961/apple-looks-to-improve-siris-script/}}. As
it is impossible to prepare answers to all the possible out-of-domain
requests, and the majority of the conversational agent's developers
cannot afford to recruit writers, a solutions is to try to take
advantage of those human dialogues that can be found in the web. An
example of a work that follows this approach is, again, the butler
Edgar Smith. In the work reported by \citet{FilipeIVA}, the previous
mentioned Filipe's corpus (Subtle), was used to answer out-of-domain
requests posed to Edgar. Reported results say that 72\% of the
out-of-domain requests asked to Edgar are now answered, and, from
these, about 65\% are considered to be appropriate answers.

Nevertheless, all these authors mention plenty of room for improvements.
Moreover, the previously mentioned corpora, made available by the
chatbot's community, were not properly explored yet. Thus, some research
questions remain to be answered:

\begin{itemize}
\item How to filter these corpora in order to \textbf{eliminate
  unwanted answers}?
\item Which techniques should be used to detect paraphrases in these
  corpora, as well as other semantic relations between requests and
  answers, in order to \textbf{organise such data}?
\item How can the appropriate answer be chosen from the set of all
  possible answers available in the corpus, in order to \textbf{allow
    some level of automatic customisation of the targeted agent}?
\item How to guarantee that a pre-defined answer \textbf{makes sense
  in the context} of a specific dialogue?
\end{itemize}

We foresee these as interesting research challenges for the next
years.
 
\section{Conclusions and Future Challenges}
\label{sec:conc}

The number of chatbots that can be found in the web increases every day.
Although the majority of their developers do not have scientific
aspirations, the fact is that, besides tools and corpora, the chatbots'
community has important know-how, which should not be neglected by
researchers targeting advances in human-computer communication.
Therefore, in this paper we presented a brief historical overview of
chatbots, and described main resources and ideas. Furthermore, we
highlighted some chatbots, which have distinguished themselves by
introducing new paradigms and/or for being Loebner prize winners.
However, it should be clear that these are only the tip of the iceberg
of the panoply of chatbots that currently exist.

We have seen that AIML and, more recently, \textsc{Chatscript} are
widely used languages that allow to code the chatbots' knowledge
sources, and that although some chatbots implement learning strategies,
scripting is still at their core. We have also seen that a personality
capable of justifying some of the chatbot's answers, the capacity of
directing a conversation and producing small talk, and the idea of
failing like a human are some of the chatbots' features that give the
illusion of intelligence.

We have also grasped that to create a chatbot, one ``only'' needs to
think about a character, and enrich its knowledge bases with possible
interactions. Even better, that work does not need to be done from
scratch as many platforms already provide pre-defined interactions,
which can be adapted according to the chatbot character. And this is the
main richness of the chatbot's community: the immense amount of
collected interactions, where the majority of them represent real human
requests.

A major future challenge is to be able to automatically use all this
information to build a credible chatbot. How to avoid contradictory
answers? How to choose appropriated  answers considering a chatbot's
character? And if we move to other sources of dialogues, like the ones
from books, theatre plays or movies subtitles, will we be able, one day,
to integrate all that information simulating real human dialogues?

\makeatletter
\def\@biblabel#1{}
\makeatother

\bibliographystyle{apalike}
\bibliography{references}

\end{document}